\documentclass[twocolumn,showpacs,preprintnumbers,amsmath,amssymb]{revtex4}

\usepackage{graphicx}
\usepackage{dcolumn}
\usepackage{bm}
\usepackage{epsfig}
\usepackage{epstopdf}

\begin{document}


\title{Generation of pure single photon wavepackets by conditional preparation
based on spontaneous parametric downconversion}

\author{Alfred B. U'Ren$^{1,2}$, Christine Silberhorn$^1$, Reinhard
Erdmann$^3$,Konrad Banaszek$^1$,\\Warren P. Grice$^4$,Ian A.
Walmsley$^1$ and Michael G. Raymer$^5$}

\affiliation{ $^1$ Clarendon Laboratory, Oxford University, Oxford,
OX1 3PU,
UK$\mbox{           }$\\
$^2$ Divisi\'{o}n de F\'{i}sica Aplicada, Centro de
Investigaci\'{o}n Cient\'{i}fica y Educaci\'{o}n Superior de
Ensenada (CICESE), Baja California, 22860, Mexico\\
$^3$ Rome Air Force Laboratory, Rome, NY, USA $\mbox{
    }$\\
$^4$Center for Engineering Science Advanced Research, Computer
Science and Mathematics Division, Oak Ridge National Laboratory, Oak
Ridge, Tennessee 37831, USA\\
$^5$Department of Physics and Oregon Center for Optics, University
of Oregon, Oregon 97403, USA}

\date{\today}

%
\newcommand{\epsfg}[2]{\centerline{\scalebox{#2}{\epsfbox{#1}}}}

\begin{abstract}
We study the conditional-preparation of single photons based on
parametric downconversion, where the detection of one photon from a
given pair heralds the existence of a single photon in the conjugate
mode. We derive conditions on the modal characteristics of the
photon pairs which ensure that the conditionally prepared single
photons are quantum mechanically pure.  We propose specific
experimental techniques which yield photon pairs ideally suited for
single photon conditional preparation.
\end{abstract}

\maketitle


\section{Introduction}

Pure single photon states are probably the most fundamental entities
in quantum optics, and constitute the starting point for many
optically-based quantum enhanced technologies. A basic requirement
for many key applications  is the ability to generate reliably pure
single photon wavepackets capable of high-visibility interference.
Single photon wavepackets may be generated from number-correlated
pairs by conditional state preparation. For example, photon pairs
produced by the process of spontaneous parametric downconversion
(PDC) in a $\chi^{(2)}$ non-linear medium allow conditional
preparation of single photons, since the detection of one photon in
the pair heralds the presence of the conjugate photon. PDC-generated
photons have been employed for both fundamental tests of quantum
mechanics and to demonstrate various quantum communication
applications, such as quantum teleportation\cite{bouwmeester97},
quantum dense coding\cite{mattle96} and quantum
cryptography\cite{gisin02}.

A recent proposal for linear optical quantum computation
(LOQC)\cite{knill01} exploits the photon bunching that occurrs in
quantum interference between pure state single photons in
conjunction with conditional state preparation  in linear optical
networks. There the detection of auxiliary photons indicates the
successful operation of a given gate. For LOQC, as for all other
applications involving optical quantum networks, high-visibility
interference between photons from multiple sources is crucial. This
necessitates, on the one hand, precise timing of the photons. This,
in turn, demands  a high probability of simultaneous generation from
synchronized distinct sources. On the other hand, precise control of
the modal structure of the generated photon pairs is essential to
guarantee the purity of the conditionally prepared photon states.
For PDC based implementations these requirements lead to the need
for a pulsed pump and for efficient sources exhibiting high
brightness and detection efficiencies. Spatial modal control of the
emitted photons, and a much increased production rate of photon
pairs may result from waveguided PDC\cite{uren03}.

A common approach to guarantee indistinguishability in optical
experiments uses strong spatial and spectral filtering.  The cost of
this, however, is a prohibitive reduction of the generation rate of
usable single photon wavepackets. Since common birefringent
phase-matching  leads to spatial-spectral correlations with
overlapping contributions, spatial filtering also causes   optical
losses for the signal photon for finite spectral filter bandwidths.
As a consequence the fidelity of the conditionally prepared signal
photons is limited due vacuum contributions. To eliminate these and
to ensure, for example, successful gate operations in LOQC
post-selecting by coincidence measurements is inevitable. But such
post-selection, in turn, hinders the realization of scalable
networks involving more than one or possibly two sources. Waveguided
PDC is one route to accomplish both, to decorrelate the spatial
degree of freedoms from spectral ones and to eliminate at the same
time the spatial correlations between the PDC photons. This implies
no need for spatial filtering, however,  conditionally prepared
single photons are in general still described by a mixed state due
to spectral correlations present in the photon pairs.  In this paper
we study the conditional state preparation of single photon
wavepackets based on PDC pumped by ultrafast pulses. We use  a
multi-mode description and propose novel engineering techniques for
eliminating correlations between unobserved variables. This ensures
that the detection of one photon from a given pair directly yields a
pure single photon state in the conjugate mode, without resorting to
filtering. The latter constitutes an important step for the further
development of quantum networks since it is expected to lead to an
increase in the probability of simultaneous generation from multiple
crystals by several orders of magnitude.

\section{Conditional state preparation of pure single photon states
in a multimode description} \label{sec:cond}

Single photon states are defined as containing a single quantum in
the photon number basis. A complete specification of the photons
requires that  all degrees of freedom associated with the quantized
electromagnetic field of the occupied optical modes must be taken
into account. Thus, the complete description of the photons involves
their polarization $\mu$, their spatial vector $\bf{r}$ or the
wavevector {$\bf{k}$}, and their frequency $\omega$ or time
dependence $t$. The dispersion relationship between $\bf{k}$ and
$\omega$ reduces one degree of freedom, such that we can restrict
our analysis to  transverse wavevectors $\bf k^\bot$. For simplicity
we also consider only one single polarization, which is appropriate
for PDC experiments aimed at conditional prepararation of single
photons. A pure single photon wavepacket then corresponds to a
coherent superposition of monochromatic plane waves:
\begin{equation}
| \phi \rangle = N \int d \omega \int d {\bf{k}} \ c(\omega,
{\bf{k}}) \ \exp[i( {\bf{k}} \cdot {\bf{r}} - \omega t) ] \
\hat{a}^{\dag}(\omega, {\bf{k}}) \ | 0 \rangle,
\end{equation}
where $N$ denotes the normalization constant and the coefficient
function $c(\omega, {\bf{k}})$ determines the modal structure of the
photon. The corresponding density operator is given by $\rho=| \phi
\rangle \langle \phi |$ and  exhibits nonzero off diagonal elements.
This is to be contrasted with a completely mixed single photon state
in frequency and momentum for which no coherence between the
different modes exists and is characterized by a density operator of
the form
\begin{equation}
\hat{\rho} = N \int d \omega \int d {\bf{k}}  |c(\omega,
{\bf{k}})|^2 \hat{a}^{\dag}(\omega, {\bf{k}})\ | 0 \rangle \langle 0
|\ \hat{a}(\omega, {\bf{k}}),
\end{equation}
which does not exhibit any off-diagonal elements.

In the process of PDC individual pump photons split into two
daughter photons in modes $\hat{a}$ and $\hat{b}$. For the purpose
of conditional state preparation the modes $\hat{a}$ and $\hat{b}$
should be distinguishable in at least one degree of freedom, which
might be polarization for type II phase-matching or the spatial
degree of freedom for non-collinear phasematching.  The interaction
Hamiltonian of the $\chi^{(2)}$-process for a strong pump field,
which is treated classically, results in  state of the approximate
form:
\begin{eqnarray}
&| \Psi \rangle = |0 \rangle | 0 \rangle + \mu
                          \iint d \omega_{s} d \omega_{i} \iint d {
\bf{k_{s}^\bot}} d
                          {\bf{k_{i}^\bot}} \nonumber \\
                         & \times F(\omega_s, {{\bf{k_s^\bot}}}}; \omega_i
                         , {\bf{k_i^\bot
                         })\
                          \hat{a}^{\dag}(\omega_s, {\bf{k_s^\bot }}) \
\hat{b}^{\dag} (\omega_i, {\bf{k_i^\bot }})\ |0 \rangle
\end{eqnarray}
where $\mu$ is a measure of the efficiency of the PDC process and
the function $F(\omega_s, {{\bf{k_s^\bot}}}; \omega_i, {\bf{k_i^\bot
})}$ represents a weighting function for the different spatial and
spectral modes present, which results from the pump envelope and
phasematching functions defined by the specific downconversion
configuration\cite{grice97}. It is assumed that states of higher
photon numbers can be ignaored.  The function $F(\omega_s,
{{\bf{k_s^\bot}}}; \omega_i, {\bf{k_i^\bot })}$ may contain
correlations between the signal and idler photons in the
spatio-temporal continuous degrees of freedom, thus yielding an
entangled two-photon pair. A useful tool for analyzing such
entanglement is that of carrying out a Schmidt decomposition
\cite{law00} of the spatio-temporal state amplitude. The Schmidt
decomposition entails expressing the two photon state in terms of
complete basis sets of orthonormal states $A_n^{\dag}| \mbox{0}
\rangle$  and $B_n^{ \dag}| \mbox{0} \rangle$ such that:
\begin{equation}
|\Psi\rangle=\sum_n\sqrt{ \lambda_n} \hat{A}_n^{\dag} \hat{B}_n^{
\dag} | \mbox{0} \rangle \label{E:schmidtdecomp}
\end{equation}
where $\sum_n \lambda_n=1$ and where the values $\lambda_n$ are
known as the Schmidt coefficients. In this description the effective
signal and idler creation operators $ \hat{A}_n^{\dag}, \
\hat{B}_n^{ \dag}$ are given in terms of the Schmidt functions
$\psi_n(\omega_s,{\bf k}_s^{\bot})$ and $\phi_n(\omega_i,{\bf
k}_i^{\bot})$ as
\begin{equation}
\hat{A}_n^{\dag}=  \ \int d {\bf k}_s^{\bot}\int d\omega_s
\psi_n(\omega_s,{\bf k}_s^{\bot}) \ \hat{a}^\dag(\omega_s,{{\bf
k}_s^\bot})|\mbox{0}\rangle
\end{equation}
and
\begin{equation}
\hat{B}_n^{\dag}= \  \int d {\bf k}_i^{\bot}\int d\omega_i
\phi_n(\omega_i,{\bf k}_i^{\bot}) \ \hat{b}^\dag(\omega_i,{{\bf
k}_i^\bot})|\mbox{0}\rangle,
\end{equation}
while the two-photon state amplitude $F({\bf k}^\bot_s ,\omega _s ;
{\bf k}^\bot_i ,\omega _i )$ can be expressed as
\begin{equation}
F( \omega _s ,{\bf k}^\bot_s;\omega_i,{\bf k}^\bot_i ) = \sum_n
{\sqrt {\lambda _n } \psi _n (} \omega _s,{\bf k}^\bot_s  )\phi _n
(\omega _i,{\bf k}^\bot_i), \label{E:schmidt}
\end{equation}
where $\omega_{s,i}$ and ${\bf k}^\bot_{s,i}$ represent the
wavelength and the transverse wavevector  of the signal and idler
photons. For the calculation of the Schmidt functions themselves the
first step is to determine the reduced density matrices of the two
sub-systems corresponding to the signal and idler photons, which are
defined by
\begin{eqnarray}
&\hat{\rho}_s(\omega_s, { \bf k^\bot_s}; \tilde{\omega}_s,\tilde{\bf
k}^\bot_s)=\iint d\omega' \ d {{\bf k^\bot}}' F(\omega_s,{\bf
k}_s^\bot,\omega',{{\bf k}^\bot}') \nonumber\\
& \times F^\ast(\tilde{\omega}_s,\tilde{\bf k}^\bot_s,\omega',{{\bf
k^\bot}}')
\end{eqnarray}
and
\begin{eqnarray}
&\rho_i(\omega_i, { \bf k}^\bot_i; \tilde{\omega}_i,\tilde{\bf
k}^\bot_i)=\iint d\omega' \ d {{\bf k^\bot}}' F(\omega',{{\bf
k^\bot}}',\omega_i,{\bf k^\bot_i}) \nonumber \\
&\times F^\ast(\omega',\tilde{\bf k}^\bot,\tilde{\omega}_i,{{\bf
k}_i^\bot }')
\end{eqnarray}
and as a second step solve the following the following integral
eigen-value equations\cite{law00}:
\begin{eqnarray}
&\iint d\omega' d { {\bf k^\bot}}'\rho_s(\omega_s,{\bf
k}_s^\bot,\omega',{ {\bf  {\bf k^\bot}}}') \psi_n(\omega',{ {\bf
{\bf k^\bot}}}')\nonumber\\&=\lambda_n \psi_n(\omega_s,{\bf
k}_s^\bot)
\end{eqnarray}
\begin{eqnarray}
&\iint d\omega' d { {\bf
  {\bf k^\bot}}}'\rho_i(\omega_i,{\bf k}_i^\bot,\omega',{ {\bf  {\bf k^\bot}}}')
\phi_n(\omega',{ {\bf  {\bf k^\bot}}}')\nonumber\\&=\lambda_n
\phi_n(\omega_i,{\bf k}_i^\bot) \label{E:SchmidtIntEq}
\end{eqnarray}
thus yielding the Schmidt functions $\psi_n$ and $\phi_n$ and the
joint eigenvalues $\lambda_n$ as Schmidt coefficients. The Schmidt
decomposition simplifies the quantum state description, by turning
the original state expressed as a quadruple integral over frequency
and transverse momentum variables into a discrete sum, typically
containing only a limited number of terms. The Schmidt functions
$\psi _n ( \omega _s,{\bf k}^\bot_s )$ and $\phi _n (\omega _i,{\bf
k}^\bot_i )$ can be thought of as the basic building blocks of
entanglement in the sense that if the signal photon is determined to
be described by a function $\psi_n$, we know with certainty that its
idler sibling is described by the corresponding function $\phi_n$.
The Schmidt modes also represent space-time localized wavepackets
\cite{titulaer65}, and may be thought of as wavefunctions of a
photon \cite{bialynicka-birula68}. The probability of extracting a
specific pair of modes $[\psi _n ( \omega _s,{\bf k}^\bot_s ), \phi
_n (\omega _i,{\bf k}^\bot_i )]$ is given by the parameter
$\lambda_n$, which is real and nonnegative. The amount of
entanglement can be conveniently quantified by the cooperativity
parameter \cite{huang93} defined in terms of the Schmidt eigenvalues
as:
\begin{equation}
K=\frac{1}{\sum\limits_n \lambda_n^2}. \label{E:K}\end{equation} The
value of $K$ gives an indication of the number of active Schmidt
mode pairs, which in turn is a measure of how much entanglement is
present in the photon pairs. A two-photon state for which the
cooperativity assumes its minimum allowed value $K=1$ represents a
state in which there is a single pair of active Schmidt mode
functions and therefore exhibits no spectral or spatial
entanglement. Another measure of the degree of non-separability is
the so-called entropy of entanglement:
\begin{equation}
S=-\sum_n \lambda_n \mbox{log}(\lambda_n) \label{E:entropy}
\end{equation}
which vanishes for a factorizable state, and increases monotonically
with the amount of entanglement present.

For the conditional state preparation of single photon states via
parametric down-conversion the idler photon is treated as a trigger,
such that in the ideal case of perfect photon number correlations, a
detection event of the trigger idler channel heralds the emission of
a photon in the signal channel. The quantum state resulting from
this conditional preparation can be obtained by modelling the
trigger detection process in terms of a projection operator and
tracing over the trigger channel photon:
\begin{equation}
\hat{\rho}_s=\mbox{Tr}_i(\hat{\rho} \hat{\Pi}). \label{E:projection}
\end{equation}
Here the  subscript $i$ denotes a partial trace acting over the
trigger idler mode, $\hat{\rho}$ is the density operator describing
the two-photon PDC state, while $\hat{\Pi}$ is the measurement
operator modelling the trigger photon detection. From the Schmidt
decomposition it can be seen easily that filtering one specific
Schmidt mode $\phi _n (\omega _i,{\bf k}^\bot_i )$  with the
projection operator $\hat{B}^\dag_n|0 \rangle   \langle 0|\hat{B}_n
$ allows conditional preparation of a pure photonic wavepacket in
the signal mode. However this procedure requires coherent
frequency-time and spatial filters, which are non-trivial to
implement.

We thus proceed to investigate the synthesis of pure signal photons
by time integrated detection of the trigger idler photon in
combination with passive spectral and spatial pinhole filters.
Correlations between the signal and idler photons generally lead to
incoherence between different frequency and spatial components upon
detection, which gives a mixed output signal state. The physical
reason for this is that a measurement of the trigger reveals some
information about the properties of the idler and hence destroys
indistinguishability.
  A spatial and
time-integrated detection of the idler state with a spectral
interference filter with a profile function $\sigma(\omega)$ and a
spatial filter of $\xi(k)$ can be modelled by the projection
operator
\begin{eqnarray}
\hat{\Pi} \ &=& \ \int dt \int d {\bf r} \iint d \omega \ d
\tilde{\omega} \iint d {\bf k} \  d \tilde{\bf k}^\bot
\sigma(\omega) \ \xi({\bf k^\bot}) \nonumber
\\ & &  \times \exp[i(
{\bf{k}} \cdot {\bf{r}} - \omega t) ]  \hat{b}^{\dag}(\omega,
{\bf{k}}) \ | 0 \rangle_i \langle 0 |_i  \hat{b}(\tilde{\omega},
\tilde{\bf{k}})\ \nonumber\\& & \times \exp[-i( \tilde{{\bf{k}}}
\cdot \tilde{{\bf{r}}} - \tilde{\omega} t) ] \
\sigma^\ast(\tilde{\omega}) \xi^\ast({\bf k^\bot}) \nonumber \\
&=& \int d  \omega  \int d {\bf k^\bot}\ |\sigma(\omega)|^2  \
|\xi({\bf k^\bot)}|^2\ \hat{b}^{\dag}(\omega, {\bf{k}}) \ | 0
\rangle_i \langle 0 |_i \nonumber \\& & \times  \hat{b}(\omega,
\bf{k}).
\end{eqnarray}
By carrying out the calculation expressed by Eq. \ref{E:projection}
using the density operator $|\Psi\rangle \langle
  \Psi|$ of the PDC state, it can be shown that the quantum state
describing the signal channel, upon registering a trigger detection
event, is given by
\begin{eqnarray}
&&\hat{\rho}_s = \int d \omega \int d {\bf k^\bot} \int d \omega_s
\int d {\bf k_s ^\bot} \int d\tilde{\omega}_s \int d {\bf
\tilde{k}_s ^\bot} \nonumber\\ &&\times|\sigma(\omega)|^2  \
|\xi({\bf k^\bot})|^2\ F(\omega_s, {{\bf{k_s^\bot}}}; \omega ,
{\bf{k^\bot}})\ \hat{a}^{\dag}(\omega_s, {\bf{k_s^\bot }}) \nonumber
\\ &&\times| 0 \rangle_s \langle 0|_s \ \hat{a}(\tilde{\omega}_s,
{\bf{\tilde{k}_s^\bot }})F^\ast(\tilde{\omega}_s,
{{\bf{\tilde{k}_s^\bot}}}; \omega , {\bf{k^\bot}})\ .
\label{E:redmatrix}
\end{eqnarray}
If we  utilize the Schmidt decomposition of Eq. \ref{E:schmidt}, the
density operator of the conditionally prepared signal photon can be
equivalently rewritten as
\begin{eqnarray}
\hat{\rho}_s &=& \sum_n \sum_m \sqrt{\lambda_n \lambda_m} \ \int d
\omega \int d {\bf k}^\bot \ |\sigma(\omega)|^2  \ |\xi({\bf
k^\bot})|^2 \nonumber\\&& \times \phi_n(\omega,{\bf k}^\bot)
\phi^*_m(\omega,{\bf k}^\bot) \hat{A}^\dag_n | 0 \rangle_s \langle
0|_s \hat{A}_m \label{E:redmatrixSchmidt}
\end{eqnarray}

Under what conditions does $\hat{\rho}_s$ represent a pure state?
Purity requires that the density operator $\rho_s$ contain a single
term, such that the signal state can be expressed in terms of a
single Schmidt mode pair. From the expressions in
Eq.~\ref{E:redmatrix} and Eq.~\ref{E:redmatrixSchmidt} two
approaches to achieve purity for the conditional state preparation
via PDC become apparent. Restricting the trigger detection bandwidth
to ideally one single spectral component, i.e. $|\sigma(\omega)|
\rightarrow \delta(\omega - \Omega) $, as well as applying strong
spatial filtering with an imaging lenses and a narrow pinhole, i. e.
$|\xi({\bf k}^\bot)| \rightarrow \delta({\bf k}^\bot - {\bf K}) $
effectively suppresses the integration over $\omega$ and ${\bf
k}^\bot$ in Eq.~\ref{E:redmatrix} and eliminates  the incoherent sum
over different frequency and spatial components, leaving the signal
photon in a monochromatic plane wave. Note, however, that this
method for purification of the prepared signal output state
necessarily implies diminishing count rates and approaches truly
pure states only in the limit of vanishing counts. The Schmidt
decompositions suggests an alternative route for the preparation of
pure  signal photon states exploiting PDC pair generation. If there
exists only one Schmidt pair present, the sums in
Eq.~\ref{E:redmatrixSchmidt} disappear and the double integral term
becomes an overall multiplicative constant. The latter implies that
the prepared photon now forms a truly quantum-mechanically pure
state. For a better understanding of this result we can relate the
Schmidt decomposition to the correlations between the generated
photons. The existence of only one Schmidt mode is actually
equivalent to the statement that the PDC two-photon state is
factorizable. Under such circumstances, the spectral and spatial
properties of signal and idler photon are independent from each
other so that the detection of the idler photon yields no
information whatsoever about the signal photon. In order to further
quantify the purity of the conditionally prepared single photons in
the signal arm we can introduce the purity parameter $p$ defined as
\begin{equation}
p=\mbox{Tr}(\hat{\rho}_s^2).
\end{equation}
Evaluating the purity $p$ in terms of the Schmidt decomposition, we
can clarify the connection between purity of conditionally prepared
\textit{single photons} and the entanglement present in PDC
\textit{photon pairs}. In the absence of filters $p=\sum_n
\lambda_n^2$, which is equivalent to $p=\frac{1}{K}$, where the
cooperativity parameter $K$ quantifies the entanglement of PDC.
Thus, entanglement between the signal and idler photon hinders pure
state preparation or, conversely, spatial and spectral separability
(which can be enhanced by appropriate filters) ensures purity for
conditional state preparation setups. This criteria delivers a
valuable tool for engineering of PDC photon pair sources optimized
for the generation of pure single photon states with without
resorting to strong filtering and illustrates the necessity of
decoupling the signal and idler photons in all degrees of freedom.

\section{Spectrally decorrelated two-photon states}

As shown in section~\ref{sec:cond}, the generation of pure single
photon wavepackets based on PDC photon pairs necessitates that all
correlations between the signal and idler photons be eliminated.
This includes correlations in all degrees of freedom including
spectral, transverse momentum and polarization. Since the full
discussion of all degrees of freedom is beyond the scope of this
paper, for the discussion following we assume that the spatial modes
of the PDC photon pairs are independently rendered decorrelated.
Such spatial decorrelation can be achieved, for example, by
waveguided PDC\cite{uren03}, in which the photons may be emitted
only into specific transversely-confined modes. In addition, some of
the techniques to be presented here designed to obtain spectral
decorrelation for bulk crystals, could be extended to the spatial
domain.

A number of techniques have been proposed to generate photon pairs
exhibiting spectral decorrelation without resorting to spectral
filtering.  Grice \textit{et al.} showed\cite{grice01} that an
avenue towards such states is the group velocity matching condition
derived by Keller \textit{et al.}\cite{keller97}. Such a technique
is extended here, showing that if an additional condition between
the pump chirp and the crystal dispersion is fulfilled,
factorizability of the joint spectral \textit{amplitude} is
guaranteed. Considering the complete joint amplitudes rather than
restricting attention to intensity correlations is important,
because correlations in the phase terms of the joint spectral
amplitude, which are not apparent in the joint spectral intensity,
can introduce correlations in the times of emission. Thus, in order
to guarantee full signal-idler decorrelation, the joint spectral
\textit{amplitude} (and therefore  the joint spectral intensity)
must be factorizable.

It has likewise been shown that a spectrally decorrelated state may
be obtained by exploiting transverse momentum in a bulk $\chi^{(2)}$
type-I crystal\cite{uren03b}.   Specifically, this is achieved using
of a focused Gaussian pump beam with a spot size fulfilling a
certain relationship with the crystal length. Walton \textit{et al.}
have recently reported a different scheme in which a transverse pump
generates counter-propagating spectrally de-correlated photon
pairs\cite{walton04}.  The essential advantage of these two
techniques is that spectral decorrelation may be obtained at any
wavelength where phasematching is possible, as opposed to the
technique based on group velocity matching which occurs for quite
restrictive frequency regimes.

In this paper we also introduce a novel technique, based on a
sequence of crystals with intermediate bi-refringent spacers. In
this scheme---although group velocities are not matched to each
other---the maximum group velocity mismatch observed can be limited
to an arbitrarily small value.  Furthermore, spectral decorrelation,
as well as a more general class of spectrally engineered photon
pairs, may be obtained at any wavelength where phasematching is
possible\footnote{Periodically poling of the
$\chi^{(2)}$-nonlinearity essentially allows to achieve
quasi-phasematching at arbitrary wavelenghts}.

\subsection{Group velocity matching} \label{sec:grvelmatch}

The general expression for the joint probability amplitude of the
two-photon state generated in the process of PDC in the ultrashort
pulsed pump regime (and assuming fixed directions of propagation)is
given by
\begin{equation}
f(\omega_s,\omega_i)=\alpha(\omega_s,\omega_i)\phi(\omega_s,\omega_i)
\end{equation}
where the pump envelope function $\alpha(\omega_s,\omega_i)$,
modelled here by a Gaussian in terms of the frequency detunings from
the central PDC frequency $\nu_\mu=\omega_\mu-\omega_0$:
\begin{equation}\label{E:PE}
\alpha(\nu_s+\nu_i)=\mbox{exp}\left[-\left (
\frac{\nu_s+\nu_i}{\sigma} \right )^2 \right ]
\end{equation}
and where the so-called phasematching function
$\phi(\omega_s,\omega_i)$ describing the optical properties of the
nonlinear crystal is expressed as\cite{grice97}:
\begin{equation}\label{E:BulkCrystal}
\phi(\omega_s,\omega_i)=\mbox{sinc}\left( \frac {L \Delta
k(\omega_s,\omega_i)}{2} \right )  e^{i \frac{L \Delta k}{2} }
\end{equation}
in terms of the phase mismatch:
\begin{equation}\Delta \label{E:phasemismatch}
k(\omega_s,\omega_i)=k_s(\omega_s)-k_i(\omega_i)-k_p(\omega_p).
\end{equation}

We will at this point restrict our attention to a
frequency-degenerate process, i.e. the Taylor expansion for each of
the two photons is assumed to be centered at the same frequency,
$\omega_0$.  We can then express the-phase-mismatch
Eq.~\ref{E:phasemismatch} as a Taylor expansion up to second order:
\begin{eqnarray}\label{E:Taylor}
L \Delta \tilde{k} (\nu_s,\nu_i) &=& L \Delta k^0 +\tau_s \nu_s
+\tau_i \nu_i \nonumber\\ &&+ \beta_s \nu_s^2+\beta_i
\nu_i^2+\beta_p \nu_s \nu_i + O(\nu^3)
\end{eqnarray}
with
\begin{equation}\label{E:k0}
\Delta k^0 = k_s(\omega_0)+k_i(\omega_0)-k_p(2 \omega_0)
\end{equation}
representing the constant term of the Taylor expansion, which must
vanish to guarantee phase-matching; $O(\nu^3)$ denotes terms of
third and higher order and ($\mu=s,i$)
\begin{align}\tau_\mu&=L\left[k_\mu'(\omega_0)-k_p'(2
\omega_0)\right] =L \left(u_\mu^{-1}-u_p^{-1} \right) \nonumber\\
\beta_\mu&=\frac{L}{2}\left[k_\mu''(\omega_0)-k_p''(2
\omega_0)\right]\nonumber\\
\beta_p&=L k_p''(2 \omega_0). \label{E:tausi}
\end{align}
In the previous expressions $'$ and $''$ denote first and second
  derivatives  with respect to frequency(evaluated at $\omega_0$ in
the case of the signal and idler wave vectors and at $2 \omega_0$ in
the case of the pump wavevector).  $u_p$ represents the pump group
velocity whereas $u_\mu \mbox{  }(\mu=s,i)$ represents the group
velocity experienced by each of the signal and idler photons.

We will use the additional approximation of expressing the sinc
function in the phasematching function as a Gaussian function
through the approximation:
\begin{equation}\label{E:PGA}
\mbox{sinc}(x)\approx \mbox{e}^{-\gamma x^2}\mbox{   with }\gamma
=.193...
\end{equation}
where the numerical value of $\gamma$ results from the condition
that both functions exhibit the same FWHM.  We will further allow
the pump field to carry a quadratic phase to account for chirp.
Under such circumstances, the joint spectral amplitude is given as:
\begin{eqnarray}
\ \check{f}(\nu_s &,& \nu_i) =   \nonumber \\
M  && \!  \mbox{exp}\left[-\left( \frac{\nu_s+\nu_i}{\sigma} \right
)^2 \right ]\mbox{exp}\left[ i
\beta_t (\nu_s+\nu_i)^2\right] \nonumber \\
\times \!\! &&\mbox{exp}\left[-\frac{\gamma}{4} (\tau_s \nu_s+\tau_i
\nu_i)^2\right]\\
  \times  \!\! &&\mbox{exp} \left[ i\frac{1}{2}(\tau_s
\nu_s+\tau_i\nu_i \nonumber    +\beta_s \nu_s^2+ \beta_i
\nu_i^2+\beta_p \nu_s \nu_i) \right] \label{E:JSAforGVM}
\end{eqnarray}
where $M$ is a normalization constant and $\beta_t$ is the GVD
dispersion term which the pump experiences prior to the crystal. We
note that the above expression contains, within the regime of the
approximations used all terms up to quadratic order in the frequency
detunings.  By expanding the exponential terms we can express the
joint spectral amplitude as
\begin{eqnarray}\label{E:JSAseparated}
&&\tilde{f}(\nu_s,\nu_i)\propto\ \nonumber\\
&& \mbox{exp}\left[ -\left(\frac{1}{\sigma^2}+\frac{\gamma}{4}
\tau_s^2 \right) \nu_s^2 \right] \mbox{exp}\left[ i \frac{\tau_s}{2}
\nu_s+i \left( \beta_t+\frac{\beta_s}{2}\nu_s^2 \right)
\right] \nonumber \\
\times&& \mbox{exp}\left[ -\left(\frac{1}{\sigma^2}+\frac{\gamma}{4}
\tau_i^2 \right) \nu_i^2 \right] \mbox{exp}\left[ i \frac{\tau_i}{2}
\nu_i+i \left(\beta_t
+\frac{\beta_i}{2}\nu_i^2 \right) \right] \nonumber \\
\times&& \mbox{exp}\left[-2
\left(\frac{1}{\sigma^2}+\frac{\gamma}{4} \tau_s \tau_i\right)\nu_s
\nu_i+i\left(2 \beta_t+\frac{\beta_p}{2} \right) \nu_s \nu_i\right].
\end{eqnarray}
Here the contributions depending on $\nu_s$ and $\nu_i$ exclusively
and those depending on both $\nu_s$ and $\nu_i$ are written as
separate factors, and phase terms linear in $\nu_\mu$ have been
dropped.  We can see from Eq.~\ref{E:JSAseparated} that any
unfactorizability resides in the mixed term containing an argument
proportional to $\nu_s \nu_i$. Furthermore, we can now easily see
that the conditions that guarantee a factorizable state (i.e. that
make the mixed term vanish) are:
\begin{equation}\label{E:ZCcond1}
\frac{4}{\sigma^2}+\gamma \tau_s \tau_i=0
\end{equation}
and
\begin{equation}\label{E:ZCcond2}
2 \beta_t+\frac{\beta_p}{2}=0.
\end{equation}

We note that the second condition [Eq.~\ref{E:ZCcond2}] refers to
phase contributions in the joint spectral amplitude, and was not
included in the earlier work by Grice \textit{et al.}\cite{grice01}.
Because such phase contributions play no role in the intensity, the
fulfilment of the first of the derived conditions
[Eq.~\ref{E:ZCcond1}] is sufficient to obtain a factorizable joint
spectral \textit{intensity}.  If both conditions
[Eqns.~\ref{E:ZCcond1} and \ref{E:ZCcond2}] are satisfied, the
two-photon state is guaranteed to be free of correlations in the
spectral and temporal domains, as indicated by a factorizable joint
spectral \textit{amplitude}. We now analyze the effects of the first
condition on the two photon state. If condition Eq.~\ref{E:ZCcond1}
is fulfilled, we may express the joint spectral intensity as:
\begin{equation}
S(\omega_s,\omega_i)=|f(\omega_s,\omega_i)|^2\propto
\mbox{exp}\left[-2 \frac{\nu_s^2}{\sigma_s^2}\right]
\mbox{exp}\left[-2 \frac{\nu_i^2}{\sigma_i^2}\right]
\label{E:FactJSI}
\end{equation}
  where the spectral widths $\sigma_s$ and $\sigma_i$ are given by:
  \begin{equation}
  \sigma_\mu=\frac{2 \sigma}{\sqrt{4+\gamma \sigma^2 \tau_\mu^2}}
\label{E:SpecWidth}
\end{equation}
with $\mu=s,i$. Thus we have confirmed that such a joint spectral
intensity exhibits the desired factorizability.  The ratio of the
larger of the two spectral widths (for signal and idler) to the
smaller one, to be referred to as the aspect ratio, (i.e. a measure
of the degree of elongation exhibited by the two-photon spectral
distribution in $ \{ \omega_s,\omega_i \}$) is given by:
\begin{equation}\label{E:AspectRatio}
r=\mbox{max}\left\{\frac{\sigma_s}{\sigma_i},\frac{\sigma_i}{\sigma_s}\right\}=\sqrt{\frac{4+\gamma
\sigma^2 \tau_1^2}{4+\gamma \sigma^2 \tau_2^2}}.
\end{equation}
where $\tau_1=\tau_s$ and $\tau_2=\tau_i$ if $\sigma_s>\sigma_i$ and
likewise $\tau_1=\tau_i$ and $\tau_2=\tau_s$ if $\sigma_i>\sigma_s$.
In what follows the aspect ratio will prove to be a useful tool in
the characterization of the two-photon state.

Let us now consider the effect of the second of the derived
conditions [Eq.~\ref{E:ZCcond2}].  We will start by analyzing the
effect of the phase terms present in the joint spectral amplitude
[Eq.~\ref{E:JSAseparated}] on the photon pair correlations. First,
the linear phases $i \tau_\mu \nu_\mu$ (with $\mu=s,i$) shift the
times of emission with respect to the pump pulse  for each of the
two single photon wavepackets. The fact that for a type-II
interaction $\tau_s \neq \tau_i$ implies that the two single-photon
wavepackets forming a given photon pair are emitted at different
mean times, where the mean emission time difference is simply
$\tau_s-\tau_i$. This  temporal walkoff can be compensated by means
of
  a relative delay between the signal and
idler photons introduced after the crystal. While the GVD terms
($\beta_s$,$\beta_i$ and $\beta_p$) have no effect on the joint
spectral \textit{intensity }of the two-photon state (in the regime
of the approximations used which neglects contributions of cubic and
higher order terms in the frequency detunings) they translate into
broadening in the temporal domain, which becomes apparent upon a
Fourier transformation of the joint spectral amplitude to obtain the
joint temporal amplitude. The quadratic phase
$i(\beta_t+\beta_\mu/2) \nu_\mu^2$ (with $\mu=s,i$) results in
uncorrelated temporal broadening of the two photon wavepackets along
the $t_s$ and $t_i$ axes. The temporal broadening is proportional to
the crystal length and is, in general,  different for each of the
two photons\footnote{In case of a degenerate type-I process
$\beta_s=\beta_i$} (since $\beta_s \neq \beta_i$). The mixed term,
proportional to $\nu_s \nu_i$,
  introduces  correlations  in the
times of emission. This means that  decorrelations of spectral
intensity are not sufficient for true single mode photon pair
generation. We note that it is the $\beta_p$ term which is
responsible for extirpating the factorizability. Therefore, if it
were possible to eliminate the effect of $\beta_p$, the two-photon
joint temporal amplitude would broaden in such a way that the
decorrelated character is maintained. Fortunately, it is possible to
compensate for the presence of $\beta_p$ simply by letting the
incoming pump field have a chirp (quadratic phase) which fulfils our
second derived factorizability condition (Eq.~\ref{E:ZCcond2}).  The
quadratic phase which the pump should carry is therefore given by:
$\beta_t=-\beta_p/4$.  We also note that it is in principle possible
to ``tune'' the degree of temporal correlation in the two-photon
state (and therefore the level of distinguishability in an
interference experiment) by using a pump field with a variable
chirp, \textit{i. e.} variable $\beta_t$.

Let us return to the first of the conditions derived for
factorizability [Eq.~\ref{E:ZCcond1}].  This condition may be
written in terms of the crystal and pump field parameters as:
\begin{equation}\label{E:ZCcond}
\frac{4}{\sigma^2}+\gamma L^2 (k'_s-k'_p)(k'_i-k'_p)=0
\end{equation}
where $\sigma$ is the pump bandwidth, $L$ is the crystal length,
$k'_p$ represents the first frequency derivative of $k_p$ evaluated
at $2 \omega_0$, $k'_\mu$ $(\mu=s,i)$ represents the first frequency
derivative of $k_\mu$ evaluated at $\omega_0$ and $\gamma \approx
.193$. For specific experimental situations, pairs of values for the
pump bandwidth $\sigma$ and the crystal length $L$ may exist such
that the above condition (Eq.~\ref{E:ZCcond}) is fulfilled. Note
that for the condition in Eq.~\ref{E:ZCcond} to be fulfilled, one of
the following must be true: $k'_s<k'_p<k'_i$ or $k'_i<k'_p<k'_s$,
\textit{i. e.} the group velocity of the pump must lie between that
of the signal and idler. None of the experiments recently reported
in the literature make use of crystals meeting the above
requirements. This should not be surprising, since it requires a
material in which one of the daughter photons has a smaller group
velocity than the pump, which necessarily has a shorter wavelength.
Alternatively, this condition may be expressed as the requirement
that $\tau_s$ and $\tau_i$ have opposite signs.

In order to get further physical insight,  consider the
phasematching condition resulting from the approximation of
neglecting second and higher order terms in the Taylor expansion
(and assuming that the constant term vanishes). In this case the
phasematching condition [see Eq.~\ref{E:phasemismatch}] becomes:
\begin{equation}
\tau_s \nu_s+\tau_i \nu_i=0
\end{equation}
The contour in $ \{ \nu_s$--$\nu_i \}$ space defined by perfect
phasematching is therefore a straight line with slope $-\tau_s /
\tau_i$. In other words the angle subtended by the $\nu_s$ axis and
the perfect phasematching contour is given by:
\begin{equation}\label{E:PMslope}
\theta_{II}=-\mbox{arctan}\left(\frac{\tau_s}{\tau_i}\right)=-\mbox{arctan}\left(\frac{k'_s-k'_p}{k'_i-k'_p}\right)
\end{equation}
where the II subscript refers to type-II phasematching. Note that
the angle $\theta_{II}$ does not depend on the crystal length; it is
solely a property of the dispersion of the material at the
particular PDC wavlength. For type-I PDC, $\tau_s=\tau_i$ so that a
similarly defined angle has the fixed value $\theta_I=-45^\circ$.
Thus, the condition that $k'_s-k'_p$ and $k'_i-k'_p$ have opposite
signs cannot be met with type-I PDC while in the case of type-II it
translates into the requirement that the slope of the perfect
phasematching contour be positive.

Let us now consider a specific case fulfilling the positive slope
requirement: the case where the slope of the perfect phasematching
contour is unity, i.e. $\theta_{PM}=45^\circ$.  It is
straightforward to show from Eq.~\ref{E:PMslope} that the condition
which guarantees such a unit slope is given by:
\begin{equation}
\frac{k'_s+k'_i}{2}=k'_p \label{E:grvelmatch}
\end{equation}
which says that the average inverse group velocities for the signal
and idler photons equals the pump inverse group velocity. This
condition, which can be alternatively expressed as $ \tau_s+\tau_i=0
$,  is  referred to as the group velocity matching condition
\cite{keller97,grice01}. We can picture this condition as the
requirement that the signal and idler photons are temporally delayed
with respect to the pump in a symmetric way, with the two PDC
photons delayed in opposite directions, by the same amount, from the
pump. Alternatively we may interpret the group velocity matching
condition Eq.~\ref{E:grvelmatch} for a factorizable two-photon state
in terms of the joint spectral intensities and identify that the
aspect ratio [see Eq.~\ref{E:AspectRatio}] evaluates to unity. The
latter implies that the factorizable joint spectral intensity
exhibits a circular shape on the $\nu_s$--$\nu_i$ plane. Such
symmetry is crucial for experiments which rely on classical
interference between signal and idler PDC modes, since it can then
fulfil the requirement of ideal signal-idler spectral mode matching
and lead to unit visibility in the conventional Hong-Ou-Mandel
experiment\cite{hong87}. Hence in this configuration the generated
PDC two photon state exhibits signal-idler mode matching  as well as
the factorizability required for ideal conditional preparation of
pure single photons, and therefore constitutes a versatile
two-photon state, useful for a broad class of
experiments\cite{uren03b}.

\subsection{Asymmetric group velocity matching}\label{Sec:Assym}

In the context exclusively of generating single pure single photon
states,  each of the crystals used must generate factorizable (but
not necessarily symmetric) two-photon states in order to guarantee
unit interference visibility in a multi-crystal experiment where
only signal or respectively idler photons are superimposed on each
other. Thus, in this context, a non-unit aspect ratio is indeed not
a limitation (as long as the elongation is ``aligned'' with the
frequency and time axes). In this section we study experimental
situations where it is in fact desirable to obtain a state
exhibiting a spectrally ``elongated'' shape and thus a high aspect
ratio.

The first  condition for frequency decorrelation
(Eq.~\ref{E:ZCcond1}) can in fact be met for several common
$\chi^{(2)}$ crystals.  For example, in the case of
periodically-poled KTP (PPKTP) (in degenerate collinear operation),
it can be met for configurations where the PDC central wavelength is
within the range $1.207 \mu m < \lambda < 2.364 \mu m$ (see also
Ref.~\cite{giovannetti02}). Within this range only the particular
wavelength of $\lambda=1.568 \mu$m  fulfills the group velocity
matching condition (Eq.~\ref{E:grvelmatch}), which yields the
unit-aspect ratio two-photon state, while the aspect ratio departs
from unity for all other wavelengths. The lower and upper bounds
correspond to two-photon states characterized by a high aspect ratio
with asymmetric joint spectral intensities, though maintaining
``alignment'' with the $\nu_s$ and $\nu_i$ axes). In
Ref.~\cite{grice01}, it was similarly established that in the case
of BBO factorizability may be obtained for wavelengths between
$1.169$nm and $1.949$nm with the group velocity matching condition
fulfilled at $1.514$nm. This spectral regio is of great importance
for optical quantum communication using optical fibers.
Unfortunately for the materials discussed above,
frequency-decorrelated PDC (and specifically unit aspect ratio
decorrelated states) occurs at wavelengths at which single-photon
detectors are not well-developed.  Although InGaAs avalanche
photodiodes have been shown to be capable of photon counting at such
longer wavelengths\cite{rarity00}, quantum efficiencies are rather
low (about $15\%$) while dark count control necessitates cryostatic
cooling. Thus, frequency decorrelated two-photon states, which occur
for available crystal materials at longer $>1\mu$m wavelengths,
present acute experimental challenges mainly in terms of detection.
In what follows we will discuss the synthesis of factorizable
two-photon states in arbitrary spectral region, with a view to
accessing an experimentally more convenient range.

We begin our analysis by re-writing the condition for
factorizability (see Eq.~\ref{E:ZCcond}) as:
\begin{equation}
\frac{4}{\sigma \tau_s}+\gamma \sigma \tau_i=0.
\end{equation}
Consider this condition in the long PDC crystal regime.  Recall that
the temporal walkoff terms $\tau_s$ and $\tau_i$ are each
proportional to the crystal length $L$. Thus, we see that if $\tau_s
>> \sigma^{-1}$ (as is the case in the limit $L\rightarrow \infty$),
the condition reduces to the simpler constraint: $\tau_i=0$: i.e.
the latter tells us that spectral decorrelation can be achieved
employing a long crystal while making one of the temporal walkoff
terms vanish.  Because in this variant of the technique the pump
group velocity is matched to that of one of the PDC photons (but not
to both) we refer to this technique as asymmetric group-velocity
matching.   Note that for a given crystal length $L$ the frequency
decorrelation condition [Eq.~\ref{E:ZCcond}] can in general be met
at most for one particular value of the pump bandwidth $\sigma$;
i.e. there is a strict one to one relationship between crystal
length $L$ and the pump bandwidth $\sigma$ required for spectral
factorizability. The latter can be an experimental limitation if for
example the pump bandwidth cannot be easily modified. In the long
crystal regime, however, the condition becomes independent of the
pump bandwidth $\sigma$. The fact that a frequency decorrelated
two-photon state may thus be synthesized irrespective of the pump
bandwidth makes such a long crystal PDC a more flexible source of
spectrally-engineered photon pairs. In such a long-crystal source,
varying the pump bandwidth merely alters the aspect ratio without
affecting the spectral decorrelation.

\begin{figure}[h]
  \epsfg{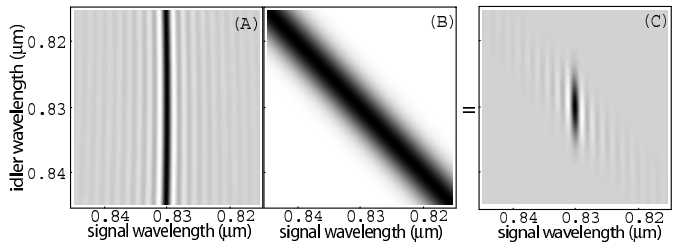}{.6} \caption{High aspect ratio
decorrelated two-photon state obtained by degenerate collinear PDC
from a 2cm-long KDP crystal exhibiting a cut angle of $67.77^\circ$.
(A) ``Vertical'' phasematching function exhibiting asymmetric group
velocity matching (B) Pump envelope function describing a pump pulse
centered at $415$nm with a spectral bandwidth of $5$nm. (C)Resulting
factorizable two-photon state.\label{Fi:ZCKDP}}
\end{figure}

Fortunately it is possible to obtain a decorrelated two-photon state
(albeit with a high aspect ratio) at wavelengths suitable for
room-temperature operated silicon-based avalanche photodiodes. These
detectors  can exhibit both $\sim60\%$ quantum efficiency as having
a manageable level of dark counts. Figure~\ref{Fi:ZCKDP} shows a
two-photon state produced by collinear degenerate type-II PDC in a
2cm-long KDP crystal cut at a phasematching angle of $68^\circ$
yielding photon pairs centered at $830$nm. Note that although the
spectral distribution is elongated, factorizability is exhibited and
hence pure conditional pure state generation (see
Sec.~\ref{sec:cond}) is ensured. It is also interesting to notice
that---in contrast to general PDC---such a state is predicted to
yield conditionally prepared photon states which are Fourier
transform limited\cite{future}. The  two-photon state features a
narrow spectrum for the ordinary-ray photon and a broader one for
the extraordinary-ray photon.  This arises  from the matching of the
group velocity on the pump photons to that of the o-ray photon, but
not to the e-ray photon.

In summary, our proposed long-crystal KDP-source yielding spectrally
factorizable photon pairs, based on asymmetric group velocity
matching, is characterized by the following properties: 1) If it is
used for conditional single photon generation the prepared photons
are described by quantum mechanically pure states. This property is
evident in the by Fourier transform limited character of the
photonic wavepackets. 2) Narrow spectral bandwidth ordinary-ray
photons allow spectral filtering to eliminate any background light
without discarding PDC photons. If the filtered mode is then used as
a trigger in a conditional source of single photons, the resulting
photon exhibits the largest spectral bandwidth allowed by the pump
bandwidth. 3) The decorrelated character is independent of the pump
bandwidth, thus eliminating an important experimental
constraint\footnote{Assuming
  the spectral bandwidth of the pump is large enough that  the
phase-matching function
  dominates over the pump envelope function}. 4) The generated state requires
a pump centered at $415$nm, easily accessible with a
frequency-doubled Titanium sapphire laser, while the produced
photons at $830$nm can be efficiently detected with Silicon-based
single photon counting modules. Even though KDP has a smaller
nonlinearity than other common $\chi^{(2)}$ materials such as BBO,
the long crystal used translates into a gain in production rate of
photon pairs with respect to typical PDC sources based on shorter
crystals. 5) KDP crystals at a $\sim70^\circ$ cut angle exhibit a
relatively small transeverse walkoff; at a pump wavelength of
$415$nm the walkoff angle is $1.15^\circ$ giving, for a $2$cm long
crystal a lateral displacement of just over $400\mu$m.  We note,
however, that used as a conditional source of single photons in
networks where signal and idler modes are not superimposed, the
walkoff exhibited is not a limitation, as the signal and idler
photons need not be indistinguishable from each other. 6) As will be
discussed in the next section, such a high aspect ratio factorizable
state additionally exhibits an interesting dispersion insensitivity
effect.

\subsection{Dispersion insensitivity in long-crystal
asymmetrically group velocity matched photon pairs}

As was discussed above, broadening of the joint spectral amplitude
results from GVD terms, which [see Eqns.~\ref{E:tausi}] are
proportional to the crystal length.  Thus, increasing the crystal
length used leads to a temporally-broader two-photon time of
emission distribution. If the mixed term phase $2
\beta_t+\frac{\beta_p}{2}$ in Eq.~\ref{E:JSAseparated} does not
vanish, such broadening eliminates the ``alignment'' with the time
axes. However, in this section we will show that, for high-aspect
ratio spectrally factorizable two-photon states,  in the
long-crystal regime  the mixed term responsible for introducing
correlations in the times of emission, can be made arbitrarily small
(even temporal broadening still occurs). In fact, an increase in
crystal length   gives a net reduction of temporal correlations.

Let us assume that the crystal length and pump bandwidth are chosen
so that the first  condition for factorizability
(Eq.~\ref{E:ZCcond1}) is satisfied.   Under these circumstances (see
Eq.~\ref{E:FactJSI}), the joint spectral amplitude may be expressed
as:
\begin{eqnarray}
f(\omega_s,\omega_i)&=&\mbox{exp}\left[-\frac{\nu_s^2}{\sigma_s^2}\right]\mbox{exp}\left[-\frac{\nu_i^2}{(r
\sigma_s)^2}\right]\nonumber\\
&&\times\mbox{exp}\left[i(\beta_s \nu_s^2+\beta_i \nu_i^2+\beta_p
\nu_s \nu_i)\right]
\end{eqnarray}
where we have expressed the idler width in terms of the aspect ratio
$r$ as $\sigma_i=r \sigma_s$ and where the GVD terms ($\beta_s$,
$\beta_i$ and $\beta_p$) were defined in Eq.~\ref{E:tausi}. Note
that because in the above expression the lack of factorizability
resides in the phase term, the joint spectral \textit{intensity} is
factorizable.   Nevertheless the presence of a mixed phase term,
proportional to $\beta_p$, translates into lack of factorizability
in the temporal domain. In order to study the temporal properties of
the wavepacket, let us carry out a Fourier transform of the above
joint spectral amplitude, to obtain the following joint temporal
intensity:
\begin{eqnarray}
\mathcal{S}(t_s,t_i)&=&|\mathcal{F}(t_s,t_i)|^2\nonumber\\
&=&\mbox{exp}\left[-\frac{2 t_s^2}{\delta
t_s^2}\right]\mbox{exp}\left[-\frac{2 t_i^2}{\delta t_i^2}\right]
\mbox{exp}\left[-2 \sigma_M^2 t_s t_i \right]
\end{eqnarray}
where the temporal wavepacket width along $t_s$ is given in terms of
the reciprocal aspect ratio $s=r^{-1}$by:
\begin{equation}\label{E:twidths}
\delta t_s=\frac{2\sqrt{2}\sqrt{2+\sigma_s^4(2 \beta_s^2+\beta_p^2
s^2+2 \beta_i^2 s^4)}}{\sigma_s \sqrt{4+\sigma_s^4 s^2 (\beta_p^2+4
\beta_i^2 s^2)}},
\end{equation}
the temporal wavepacket width along $t_i$ is given by:
\begin{equation}\label{E:twidthi}
\delta t_i=\frac{2\sqrt{2}\sqrt{2+\sigma_s^4(2 \beta_s^2+\beta_p^2
s^2+2 \beta_i^2 s^4)}}{s \sigma_s \sqrt{4+\sigma_s^4 (4 \beta_s^2+4
\beta_p^2 s^2)}},
\end{equation}
and
\begin{equation}\label{E:sigmaM}
\sigma_M^2 =\frac{ \sigma_s^6 s^2 \beta_p(\beta_s+\beta_i
s^2)}{2(2+\sigma_s^4(2 \beta_s^2+\beta_p^2 s^2+2 \beta_i^2 s^4))}
\end{equation}
represents the mixed term coefficient. Specializing to a high aspect
ratio spectrally factorizable state such as one produced by the long
KDP cyrstal source discussed in the last section, the temporal
walkoff between pump and the ordinary-ray photon $\tau_o$ vanishes.
Thus, according to Eq.~\ref{E:SpecWidth} the ordinary-ray spectral
width is then given by $\sigma_s=\sigma$ where $\sigma$ represents
the pump bandwidth. Likewise, in the long-crystal limit, according
to Eq.~\ref{E:SpecWidth} the idler spectral width is given by
$1/(\sqrt{\gamma}{\tau_e})$. This leads to the following expression
for the reciprocal aspect ratio:
\begin{equation}
s=\frac{2}{\sqrt{\gamma}\sigma (k'_s-k'_p)L}.
\end{equation}

Note, from the above expression, that the reciprocal aspect ratio
$s$ is inversely proportional to the crystal length $L$.  This fact,
together with the linear dependence of each of the GVD terms
($\beta_s$,$\beta_i$ and $\beta_p$) on the crystal length $L$ can be
used to express the mixed term coefficient $\sigma_M^2$ in terms of
its overall length dependence. We thus obtain:
\begin{eqnarray}\label{E:sigmaMsimp}
\sigma_M^2&\approx&\frac{\sigma^6 s^2 \beta_p \beta_s}{4(1+\sigma^4
\beta_s^2)}\nonumber\\&&=\frac{\sigma^4 k_p'(k_s'-k_p')}{4 \gamma
(k_i'-k_p')^2\left [1+\sigma_p^4 L^2(k_s'-k_p')^2 \right ]}.
\end{eqnarray}
The net crystal length dependence of $\sigma_M^2$ in the
long-crystal regime is inverse quadratic. Hence increasing the
crystal length  does not lead to an increase in magnitude of the
mixed term, but  actually reduces the correlations. Recall that for
a factorizable high aspect ratio two photon state, the pump
bandwidth $\sigma$ and crystal length $L$ do not have to be
specified jointly \textit{i.e.} there is no  relationship that
should be fulfilled between them to ensure factorizability. Together
with the dispersion insensitivity, the latter means that the crystal
can be made arbitrarily long without introducing temporal
correlations.  The main remaining considerations using a long
crystal are excessive temporal broadening and absorption of PDC
photons in the crystal.  We note that if the pump field is chirped,
$\beta_p$ should be substituted by $2 \beta_t+ \frac{\beta_p}{2}$ in
the expressions for the mixed term, Eqns.~\ref{E:sigmaM} and
\ref{E:sigmaMsimp} (where $\beta_t$ is introduced in
Eq.~\ref{E:JSAforGVM} represents the quadratic phase carried by the
pump field).  The presence of chirp in the pump pulse can be treated
similarly: upon increasing the crystal length $L$, the magnitude of
the mixed term decreases. Thus, we conclude that (within the regime
of validity of the approximations used) the source presented here is
insensitive to those dispersive effects leading to the appearance of
temporal correlations, either introduced in the PDC crystal itself
or indeed prior to the crystal.

\begin{figure}[h]
\label{Fi:KDPJSIandJTI} \epsfg{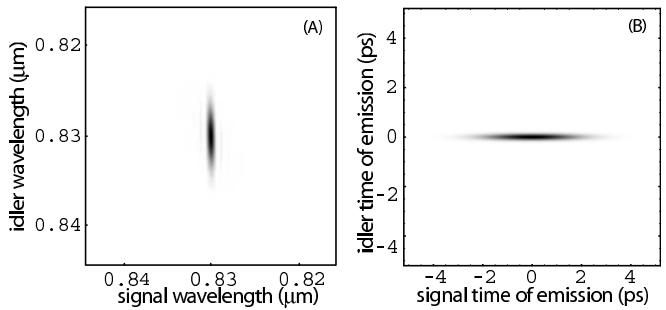}{.6} \caption{High
aspect ratio two-photon state obtained by degenerate collinear PDC
from a 2cm-long KDP crystal exhibiting space time decorrelation (A)
Factorable joint \textit{spectral} intensity (B) Factorable joint
\textit{temporal} intensity.}
\end{figure}

We can understand the dispersion insensitivity effect described
above by realizing that dispersion effects in general require not
only the presence of a dispersive phase but also broadband light.
Engineering the joint spectral intensity by making it nearly
monochromatic along either the signal or the idler axis, means that
the resulting narrowband photon will experience  limited dispersion.
Furthermore, the mixed term (leading to temporal correlations) is
also reduced by the monochromaticity of one of the photons.
Fig.~\ref{Fi:KDPJSIandJTI} shows for the two-photon state produced
by collinear degenerate type-II PDC in a 2cm-long KDP crystal cut at
a phasematching angle of $68^\circ$ yielding photon pairs centered
at $830$nm.  The joint \textit{spectral} intensity is shown in
Fig.~\ref{Fi:KDPJSIandJTI}(A) while the joint \textit{temporal}
intensity in Fig.~\ref{Fi:KDPJSIandJTI}(B). Because the parameters
of the source fulfil Eq.~\ref{E:ZCcond}, the joint spectral
intensity is factorizable. Furthermore, the joint temporal intensity
is also factorizable, for arbitrarily long crystals. Thus, the
two-photon state produced by such a two-photon source is free from
spectral (temporal) correlations, making it an ideal source of
conditionally prepared pure and Fourier-transform limited single
photon wavepackets.

\section{Two photon state engineering via crystal sequences with
bi-refringent compensators}

As has been shown in this paper, group velocity matching is a
powerful technique which enables the engineering of two photon
states with desirable properties, tailored for specific
applications. Nevertheless, group velocity matching crucially
depends on the dispersion exhibited by the pump, signal and idler
fields in the nonlinear crystal used, and tends to occur only in
specific wavelength ranges.  In this section we propose and analyse
a technique in which the group velocity mismatch can be controlled
at arbitrary wavelenghts. This comes
  at the cost, however, of an increased source
complexity. In this scheme, a \textit{sequence} of nonlinear
crystals is used interspersed with birefringent spacers exhibiting a
dispersion such that the group velocity mismatch introduced by the
crystal is compensated for by that in the spacer. This scheme can be
thought of as analogous to band gap engineering using photonic
crystals, except that group velocity, rather than phase velocity, is
compensated.

\begin{figure}[h]
\epsfg{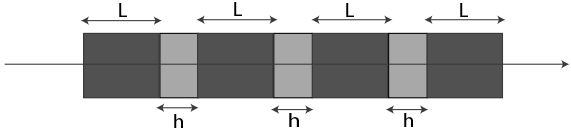}{.4} \caption{This figure shows a schematic of
the proposed crystal sequence with intermediate bi-refringent
spacers.  Each crystal has a length $L$ while each spacer has a
length $h$.\label{Fig:XtalSeq}}
\end{figure}
Consider the experimental arrangement shown in
Fig.~\ref{Fig:XtalSeq} consisting of $N$ identical $\chi^{(2)}$
crystals and $N-1$ linear $\xi^{(1)}$ spacers.  Each of the crystals
is assumed to be cut and oriented for degenerate collinear type-II
PDC while the compensators are assumed not to exhibit a $\chi^{(2)}$
nonlinearity. It is further assumed that each crystal has length $L$
while each spacer has length $h$. The phase mismatch in each of the
crystals is given by
\begin{equation}
\Delta k=k_p-k_s-k_i
\end{equation}
where $k_\mu$ (with $\mu=p,s,i$) denotes the wavenumber for each of
the three fields taking into account dispersion in the crystal. The
phase mismatch introduced by each of the spacers is equivalently
given by:
\begin{equation}
\Delta \kappa=\kappa_p-\kappa_s-\kappa_i.
\end{equation}
where $\kappa_\mu$ (with $\mu=p,s,i$) now represents the wavenumber
for each of the three fields taking into account dispersion in the
birefringent spacer. For an assembly of $N$ crystals and $N-1$
spacers  the overall phasematching function can then be calculated
as
\begin{eqnarray}\label{E:FullPMF}
\phi_N(\Delta k, \Delta \kappa)&=&\sum\limits_{m=0}^{N-1}
\mbox{e}^{i m (L \Delta k+h \Delta \kappa)}
\mbox{sinc}\left[\frac{L}{2} \Delta k\right] \nonumber \\&&
=\mbox{e}^{\frac{i(N-1)\Phi}{2}}\frac{\mbox{sin}(\frac{N\Phi}{2})}{\mbox{sin}(\frac{\Phi}{2})}\mbox{sinc}\left[\frac{
L \Delta K}{2}\right]
\end{eqnarray}
where we defined the quantity $\Phi$ as
\begin{equation}
\Phi=L \Delta k+h \Delta \kappa.
\end{equation}
Hence, apart from an overall phase factor the crystal assembly
phasematching function is composed  of the product of two distinct
functions: one corresponds to the phasematching function of a single
crystal and the second factor incorporates the combined effect of
the crystal and spacer dispersion. In order to carry out more
explicit calculations, it is helpful to write down the crystal and
spacer phasemismatch as a Taylor expansion [similar to that in
Eq.~\ref{E:Taylor}, however here we omit all terms of orders higher
than $O(\nu^2)$ ]. Thus, we obtain, in terms of the frequency
detunings $\nu_\mu=\omega_\mu-\omega_0$ (with $\mu=s,i$):
\begin{equation}\label{E:phi}
\Phi=L \Delta k^{(0)}+h \Delta \kappa^{(0)} + T_s \nu_s+T_i \nu_i
\end{equation}
where $L \Delta k^{(0)}$ and $h \Delta \kappa^{(0)}$ denote the
constant terms of the Taylor expansions for the crystal and spacer
phase mismatch terms and
\begin{equation}\label{E:Tfirstorder}
T_\mu=\left[k_p'(2\omega_0)-k_\mu'(\omega_0)\right]L
+\left[\kappa_p'(2\omega_0)-\kappa_\mu'(\omega_0)\right]h
\end{equation}
with $\mu=s,i$ constitutes the first-order coefficients of the
expansion. The term $\Delta k^{(0)}$ vanishes under the assumption
that each of the crystals is aligned such that phasematching is
attained.

\begin{figure}[h]
\label{Fig:upsilon} \epsfg{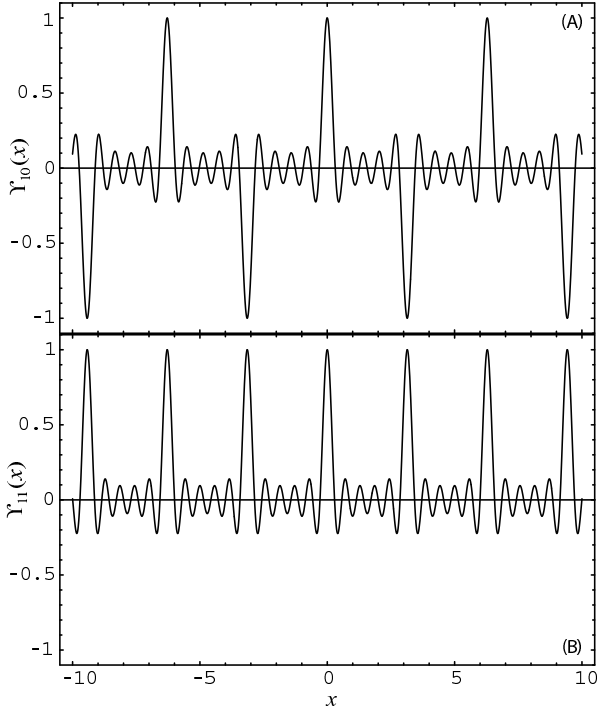}{.4} \caption{Plots of the
$\Upsilon_N(x)$ function showing different behavior for even and odd
N; while for even N, peaks alternate with valleys (separated by
$\pi$) for odd N, there are no valleys.   The peaks and valleys
become narrower with increasing N.   Such a function describes the
phasematching contribution from the crystal-birefringent spacer
sequence.}
\end{figure}

We can now analyze the influence of the proposed crystal and
birefringent compensator assembly on the joint spectral correlations
and introduce a new function
\begin{equation}
\Upsilon_N(x)=\frac{1}{N}\frac{\sin(Nx)}{\sin(x)},
\end{equation}
which describes the modifications to the spectral structure of the
phasematching function [see Eq.~\ref{E:FullPMF}]. As depicted in
Fig.~\ref{Fig:upsilon}  the function $\Upsilon_N(x)$ exhibits for
large values of $N$ a periodic structure of narrow peaks, which are
separated by $2 \pi$. While for odd values of $N$, all peaks are
positive, for even values of $N$ the peaks alternate between
positive and negative values.  The width of the peaks diminishes
with increasing $N$.
We may, thus, write the resulting phasematching function (where the
phase term in Eq.~\ref{E:FullPMF} is to be neglected) for the
crystal assembly as
\begin{equation}
\phi_N(\nu_s,\nu_i)=\phi(\nu_s,\nu_i)\chi(\nu_s,\nu_i;N)
\end{equation}
where $\phi(\nu_s,\nu_i)$ represents the \textit{single-crystal}
phasematching function and $\chi(\nu_s,\nu_i;N)$ the crystal
assembly contribution. If we express the phasemismatch as Taylor
series considering only terms up to the first order, we obtain for
the phase matching function the contributions
\begin{equation}
\chi(\nu_s,\nu_i;N)= \Upsilon_N\left(\frac{1}{2}\left[h \Delta
\kappa^{(0)} +T_s \nu_s+T_i \nu_i\right]\right).
\end{equation}
and
\begin{equation}
\phi(\nu_s,\nu_i)=\mbox{sinc}\left[\frac{L}{2}(\tau_s \nu_s+ \tau_i
\nu_i) \right]
\end{equation}
where the $\tau_\mu$ values are similar to $T_\mu$ [with $\mu=s,i$]
(see Eq.~\ref{E:Tfirstorder}), except for the absence of the spacer
contribution:
\begin{equation}
\tau_\mu=\left[k_p'(2\omega_c)-k_\mu'(\omega_c)\right]L.
\end{equation}

Let us now study the increased flexibility introduced into the
two-photon state by such an assembly contribution to the
phasematching function.  We start our analysis by noting that the
contours of the $\chi(\nu_s,\nu_i;N)$ function are straight lines
oriented with a slope given by:
\begin{equation}
\tan\theta=-\frac{T_s}{T_i}=-\frac{(k_p'-k_i')L+(\kappa_p'-\kappa_i')h}{(k_p'-k_s')L+(\kappa_p'-\kappa_s')h}
\end{equation}
Unlike for the case of a single crystal, where a change in crystal
thickness merely results in an alteration of the spectral width (and
further modifications to the phasematching function cannot be
attained except using different spectral regions or using different
crystal materials), in the case of our crystal sequence source, the
orientation of the spectral phasematching structure can be adjusted
at will by varying the crystal to spacer thicknesses ratio $L/h$.
The latter represents a very important added flexibility as it means
that in principle an arbitrary orientation for the phasematching
function orientation can be obtained.

  To illustrate the power of such engineering of PDC
we will discuss in the following the specific case of synthesizing a
phasematching function with unit slope contours. As seen in section
\ref{sec:grvelmatch} (see also \cite{grice01}) such unit slope can
result in spectrally decorrelated two-photon states. For the crystal
assembly the inverse group velocities $k_\mu'$, $\kappa_\mu'$ (with
$\mu=i,j$) evaluated at $\omega_0$ and $k_p'$, $\kappa_p'$ evaluated
at $2\omega_0$,  and the lengths $L$ and $h$ should satisfy the
condition
\begin{equation}\label{E:genGVM}
(k_s'+k_i'-2k_p')L+(\kappa_s'+\kappa_i'-2\kappa_p')h=0
\end{equation}
  Eq.~\ref{E:genGVM} closely resembles the group velocity
matching condition [see Eq.~\ref{E:grvelmatch}]. In fact, this
condition tells us that unit-slope phasematching contours require
that a generalized group velocity condition be fulfilled: the
weighted average (with crystal and spacer lengths as weighting
factors) of the group velocity mismatch in the crystal and the
spacer must vanish. Note that for this condition to be attainable,
the spacer must exhibit a group velocity mismatch with the opposite
sign as that of the crystal itself. A second condition which must be
fulfilled relates to the constant phase term $h \Delta \kappa^{(0)}$
[see Eq.~\ref{E:phi}].  If such a term differs from a multiple of
$2\pi$, the maxima of the $\Upsilon_N(x)$ function can shift away
from degeneracy (indeed by tuning such constant phase term, the
resulting modes shift on the $\nu_s$--$\nu_i$ plane).  Thus, a
second condition on the spacer length which must be met is as
follows:
\begin{equation}\label{E:condonh}
h=\frac{2\pi m}{\Delta \kappa^{(0)}}
\end{equation}
with $m$ being an integer number.  This tells us that the spacer
length should be given as an integer multiple of the quantity $2 \pi
/ \Delta \kappa^{(0)}$, which limits the minimum allowed spacer
thickness. Thus, if the conditions expressed by Eqns.~\ref{E:genGVM}
and \ref{E:condonh} are fulfilled, the crystal assembly
phasematching function can be written as:
\begin{equation}
\chi_N(\nu_s,\nu_i)= \Upsilon_N(T_{-}[\nu_s-\nu_i])
\end{equation}
where:
\begin{equation}\label{E:Tminus}
T_{-}=\frac{1}{2}\left([k_s'(\omega_0)-k_i'(\omega_0)]L+[\kappa_s'(\omega_0)-\kappa_i'(\omega_0)]h\right).
\end{equation}
Such a function consists of ``ridges'' (alternated with ``trenches''
for even N) described by unit-slope contours with a FWHM (along the
$\nu_s-\nu_i$ direction) given, in the limit of large N, by
\begin{equation}
\delta \lambda=\frac{\sqrt{2} \lambda_0^2 \gamma_2}{\pi c N T_{-}}
\end{equation}
with $\gamma_2=1.39156..$ (this numerical value results from
calculating the width of the $\Upsilon_n(x)$ function) and
$\lambda_0$ denotes the central PDC wavelength. The maxima of
different ``ridges'' result from the argument of the $\Upsilon_N$
yielding multiples of $2 \pi$.   In the case studied here where
group velocity matching is attained [see Eq.~\ref{E:genGVM}], the
spectral separation between subsequent ridges along the $\nu-\nu_i$
direction can be shown to be given by
\begin{equation}\label{E:DeltaLambda}
\Delta \lambda= \frac{\lambda_0^2}{\sqrt{2} c T_{-}}.
\end{equation}
Note that the generation of spectrally decorrelated state actually
necessitates a source with a single ``ridge''; multiple ridges
yields a joint spectral intensity with multiple peaks in
$\nu_s$--$\nu_i$ space which precludes spectral factorizability.
Because a single ridge is not attainable with our crystal assembly
source, we aim to let the separation between subsequent ridges be as
as large as possible so that the portion of the quantum state
exhibiting the desired behavior can be isolated, e.g. by weak
spectral filtering. We would like to emphasize that such spectral
separation between subsequent ridges $\Delta \nu$ is inversely
proportional to the parameter $T_{-}$, which in turn is proportional
to the two thicknesses $L$ and $h$. This leads to the manufacturing
constraint that the individual crystal and spacer lengths should be
as short as possible.  As will be shown below this technique can
indeed lead to a source of a decorrelated state with experimentally
feasible crystal and spacer thicknesses.

Altogether the resulting joint spectral amplitude for the two-photon
state produced by such a crystal assembly is therfore given by:
\begin{equation}
f(\nu_s,\nu_i)=\alpha(\nu_s+\nu_i)\phi(\nu_s,\nu_i)\chi(\nu_s,\nu_i;N)
\end{equation}
where $\alpha(\nu_s+\nu_i)$ represents the pump envelope function,
which is to be modelled as a Gaussian as in Eq.~\ref{E:PE}.  Our
approach is to let the bandwidth of the single crystal phasematching
function $\phi(\nu_s,\nu_i)$ be much larger than that of the
function $\chi(\omega_s,\omega_i;N)$, resulting from the crystal
assembly.  Thus, the spectral structure imposed on the phasematching
function by the crystal assembly dominates over that of a single
crystal. Given that the bandwidth of individual ridges of the
function $\chi(\nu_s,\nu_i;n)$ depends inversely on $N$ (while that
of the function $\phi(\nu_s,\nu_i)$ does not depend on $N$), it is
possible simply by using a large enough number of crystals to ensure
that this condition on the widths of these two functions is
fulfilled.   Let us note that for the case of interest when the
generalized group velocity matching condition [see
Eq.~\ref{E:genGVM}] is satisfied, the orientations of
$\chi(\nu_s,\nu_i;N)$ and that of the pump envelope function are
orthogonal to each other (along the $\nu_s+\nu_i$ and $\nu_s-\nu_i$
directions respectively). This leads to the important conclusion
that by choosing the pump bandwidth appropriately, a factorizable
two-photon state can be obtained exhibiting nearly-circular
contours.   The value for the pump bandwidth $\sigma$ which
guarantees such factorizable behavior can be expressed, in terms of
the number of crystal segments $N$ and the parameter $T_{-}$ (see
Eq.~\ref{E:Tminus}) as:
\begin{equation}\label{E:condonsigma}
\sigma=\frac{2\sqrt{2}}{\sqrt{\mbox{ln}(2)}} \frac{\gamma_2}{N
T_{-}}.
\end{equation}

In summary, the proposed strategy for synthesizing a decorrelated
two-photon state is as follows:  i) The crystal and spacer materials
should be chosen such that the group velocity mismatch for the
crystal and spacer material have opposite signs. ii) The crystal and
spacer thicknesses $L$ and $h$ should be chosen so that the
generalized group velocity matching condition [see
Eq.~\ref{E:genGVM}] is fulfilled \textit{and} such that the constant
phase term $\Delta \kappa^{(0)}$ vanishes [see Eq.~\ref{E:condonh}].
Additionally, the crystal and spacer should be thin enough so that
the spectral separation between modes of the $\chi(\nu_s,\nu_i;N)$
function is large enough to ensure that a single resulting mode may
be isolated. iii) The number of crystal segments $N$ should be made
large enough so that the phasematching bandwidth of a single crystal
is much larger than that of the assembly phasematching function
$\chi(\nu_s,\nu_i;N)$. iv)  The pump should be prepared so that it
exhibits a bandwidth $\sigma$ which fulfils Eq.~\ref{E:condonsigma},
thus guaranteeing that the pump envelope and phasematching functions
are such that they yield a factorizable two-photon state.

\begin{figure}[h]
  \epsfg{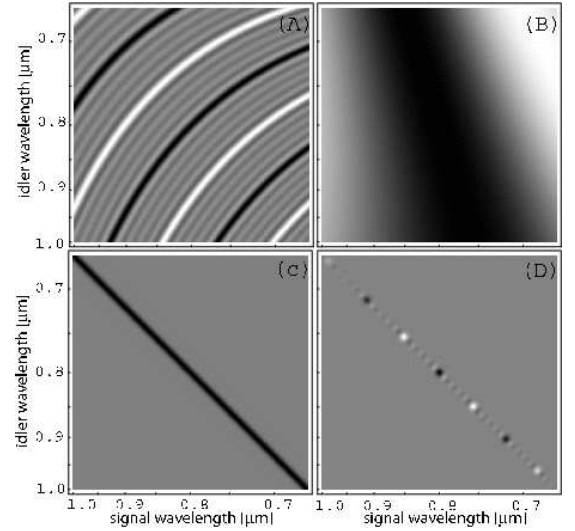}{.45} \caption{(A)
Assembly contribution to the phasematching function for the case of
10 BBO crystals with a cut angle of $\theta_c=42.35^\circ$ (each
with a thickness of $48.85\mu$m) alternated with $9$ calcite spacers
(each with a thickness of $58.83\mu$m).  Black areas indicate
positive values while white areas indicate negative values. (B)
Phasematching function for single BBO crystal, with above
characteristics. (C) Pump envelope function for pump field centered
at 400nm with a bandwidth of $1.48$nm. (D) Resulting joint spectral
amplitude function.\label{Fig:seqPMF}}
\end{figure}

We now study a specific example involving an assembly of BBO
crystals and calcite spacers.  Each of the crystals is characterized
by a cut angle of $\theta_c=42.35^\circ$ yielding type-II collinear
degenerate phasematching at a central PDC wavelength of
$\lambda_c=800$nm.  We remark that BBO and calcite exhibit the
required opposite sign group velocity mismatch; for BBO:
$2k_p'-k_s'-k_i'=3.535 \times 10^{-4} \mbox{ps}/\mu\mbox{m}$ whereas
for calcite: $2k_p'-k_s'-k_i'=-2.936 \times 10^{-4}
\mbox{ps}/\mu\mbox{m}$.  The generalized group velocity matching
condition [see Eq.~\ref{E:genGVM}] tells us that the ratio of the
calcite spacer thickness to the BBO crystal thickness should be
$h/L=1.204$.  This yields a value for the minimum spacer thickness
$2 \pi/ \Delta\kappa^{(0)}=5.88 \mu$m; we choose a value for the
integer $m$ [see Eq.~\ref{E:condonh}] of $10$, thus yielding a
spacer thickness $h=58.83\mu$m and a BBO crystal thickness
$L=48.85\mu$m.  These thicknesses yield a spacing between assembly
phasematching modes (along the $\nu_s-\nu_i$ direction) of $\Delta
\lambda=67.05$nm, corresponding to a separation along the $\nu_s$ or
$\nu_i$ axes of $47.41$nm.  The latter means that over a bandwidth
of $95$nm around the central PDC wavelength (of $800$nm) there is a
single crystal-spacer assembly phasematching mode present.
Fig.~\ref{Fig:seqPMF}(A) shows a plot  of the function
$\chi(\nu_s,\nu_i;10)$ for the above parameters,\textit{ i.e.}
assuming that the assembly contains 10 BBO crystals and 9 calcite
spacers. Note that near degeneracy (at $\lambda=800nm$), the slope
of the contours defining the function is indeed unity.
Fig.~\ref{Fig:seqPMF}(B) shows a plot of the phasematching function
$\phi(\nu_s,\nu_i)$ for a single BBO crystal (with the above
characteristics) making it clear that the spectral bandwidth
exhibited is much larger than that of the assembly phasematching
modes; the latter implies that the overall phasematching spectral
structure is  dominated by the assembly rather than the single
crystal contribution. Fig.~\ref{Fig:seqPMF}(C) shows the pump
envelope function corresponding to a pump field with a center
wavelength of $400$nm and a spectral FWHM of $1.48$nm, as required
by Eq.~\ref{E:condonsigma} to yield a factorizable state.
Fig.~\ref{Fig:seqPMF}(D) shows the resulting joint spectral
amplitude for our crystal assembly source. Due to the periodic
character of the $\Upsilon_N$ function there are multiple modes,
separated along each frequency axis by $47.4nm$. Nevertheless, weak
spectral filtering can isolate the central mode (centered at
$\lambda_s=\lambda_i=800nm$). Fig.~\ref{Fig:seqPMFJSI} shows a the
joint spectral intensity corresponding to this central mode, clearly
exhibiting spectral decorrelation.

\begin{figure}[h]
\label{Fig:seqPMFJSI} \epsfg{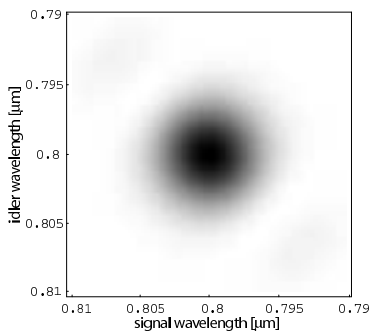}{.35} \caption{This figure
shows a the joint spectral intensity of the two-photon state
produced by a sequence of 10 BBO crystals and 9 calcite spacers, as
discussed in the text.  The plot shows the region of the main
degenerate peak near
$\lambda_s=\lambda_i=800$nm.\label{Fig:seqPMFJSI}}
\end{figure}

We have  presented a novel technique which can yield factorizable
states, based on group velocity matching compensation using a
sequence of crystal segments alternated with bi-refringent spacers.
The essential advantage of this approach is that it becomes in
principle possible to synthesize a wide class of two-photon states
simply by varying the spacer to crystal spacer thickness ratio,
while choosing the crystal and spacer materials so that they obey
certain relative dispersive properties. This scheme does not
altogether eliminate the need for filtering; however, by reducing
the crystal and spacer thicknesses $L$ and $h$ (and thus increasing
$\Delta \lambda$; see Eq.~\ref{E:DeltaLambda}) it becomes possible
to increase the separation between modes, ideally to the point where
in practice a single mode is present. In the example shown above,
the thicknesses were chosen in the region of $50\mu$m, since thinner
segments may be difficult to manufacture and to handle. However, it
is possible that by doping, ion exchange or some other process a
single crystal may be modified in certain regions in such a manner
that the group velocity mismatch changes sign; the modified regions
would then play the role of the spacers, however in a single
monolithic structure. With this approach it may be possible to take
advantage of the smallest thicknesses allowed (with a value of $m=1$
in Eq.~\ref{E:condonh}), and thus eliminate the challenges likely to
be faced in assembling a large number of thin crystals.

\section{Conclusions}

We have studied single photon conditional preparation based on
photon pairs generated by the process of parametric downconversion.
By modelling the detection as an appropriately defined projection
operator we showed that, in general, the quantum state of the
prepared state is described as a statistical mixture of the modes
defined by the Schmidt decomposition of the PDC two-photon state.
Furthermore, it was determined that a two photon state lacking
correlations between the signal and idler photons, in all degrees of
freedom, is a basic requirement for ideal single photon preparation,
i.e. yielding quantum-mechanically pure single photons.   Such pure
single-photon wavepackets are of fundamental importance in
experiments relying on quantum interference of single photons from
distinct sources, as is the case in the recently proposed scheme for
quantum computation with linear optics\cite{knill01}.

Assuming that spatial correlations can be independently eliminated
(for example by the use of waveguided PDC\cite{uren03}), we have
discussed two techniques resulting in spectrally factorizable
two-photon states.   Firstly, it was shown that if an additional
condition on the two photon state is fulfilled, the technique
reported by Grice \textit{et al.}\cite{grice01} can yield a state in
which both the joint \textit{temporal} intensity, and the joint
\textit{spectral} intensity are factorable. The latter is crucial to
guarantee full factorizability in two photon states designed for
conditional preparation of single photons. Secondly, we have
introduced a novel technique in which the group velocity mismatch
between the pump and downconverted photons, responsible for the
mixedness of the prepared single photons, can be controlled by using
a sequence of crystals alternated with birefringent spacers
exhibiting a dispersion which compensates that of the crystal in a
specific manner.  This represents a powerful technique in which
spectrally decorrelated states, as well as a more general class of
spectrally engineered two photon states, can be obtained simply by
varying the relative thicknesses of the crystal and spacer used.
These techniques described may provide useful tools for practical
implementations of novel quantum-enhanced technologies, such as
linear optical quantum computation.

\begin{acknowledgements}
This work was funded in part by the EPSRC (CS, IAW), the US National
Sciences Foundation ITR Program (MGR, IAW, KB) the US Air Force Rome
Laboratory (RE), the US Army Research Office through the MURI
program (AU) and ARDA (WG, IAW). We gratefully acknowledge this
support.
\end{acknowledgements}


\begin{thebibliography}{99}

\bibitem{bouwmeester97} D. Bouwmeester \textit{et al.}, 1997,  Nature
\textbf{390}, 575;
  D. Boschi \textit{et al.}, 1998, Phys. Rev. Lett. \textbf{80},
1121.

\bibitem{mattle96}K. Mattle, H. Weinfurter, P.G. Kwiat and
A.Zeilinger, 1996, Phys. Rev. Lett. \textbf{76}, 4656–4659.

\bibitem{gisin02} See for example review: N. Gisin, G.G. Ribordy, W.
Tittel and H. Zbinden, 2002, Rev. of Mod. Phys. \textbf{74}, 145

\bibitem{knill01} E. Knill, R. LaFlamme and G.J. Milburn, 2001,
Nature \textbf{409},  46; T.C. Ralph, A.G. White, W.J. Munro and
G.J. Milburn, 2001, Phys. Rev. A \textbf{65}, 012314; T.B. Pittman,
M.J. Fitch, B.C. Jacobs and J.D. Franson, 2003, quant-ph/0303095.

\bibitem{uren03} A.B. U'Ren, C. Silberhorn, K. Banaszek and I.A.
Walmsley, 2004, Phys. Rev. Lett. \textbf{93}, 093601

\bibitem{titulaer65} U.M. Titulaer, R.J. Glauber, 1966, Phys. Rev.
\textbf{145}, 1041.

\bibitem{bialynicka-birula68} Z. Bialynicka-Birula, 1968, Phys.
Rev. \textbf{173}, 1207.

\bibitem{grice97} W.P. Grice and I.A. Walmsley, 1997, Phys. Rev. A,
\textbf{56}, 1627.

\bibitem{law00} C.K. Law, I.A. Walmsley and J. H. Eberly, 2000, Phys. Rev.
Lett., \textbf{84}, 5304.

\bibitem{huang93} H. Huang and J.H. Eberly, 1993, J. Mod.
Opt.,\textbf{40},915

\bibitem{grice01} W.P. Grice, A.B. U'Ren and I.A. Walmsley, 2001, Phys. Rev. A
\textbf{64}, 063815

\bibitem{keller97}T.E. Keller and M.H. Rubin, 1997, Phys. Rev. A,
\textbf{56}, 1534

\bibitem{hong87}C.K. Hong, Z.Y. Ou and L. Mandel, 1987, Phys. Rev.
Lett., \textbf{59}, 2044

\bibitem{uren03b} A.B. U'Ren, K. Banaszek and I.A. Walmsley, 2003,
Quantum Information and Computation \textbf{3}, 480.

\bibitem{walton04} Z.D. Walton, A.V. Sergienko, B.E.A. Saleh and M.C.
Teich, 2004, Phys. Rev. A \textbf{70}, 052317

\bibitem{giovannetti02} V. Giovannetti and L. Maccone and J. H.
Shapiro and F. N. C. Wong, 2002, Phys. Rev. A, \textbf{66}, 043813.

\bibitem{rarity00}J. Rarity and T. Wall and K. Ridley and P. Owens
and P. Tapster, 2000, Appl. Opt., \textbf{39}, 6746.

\bibitem{future}A.B. U'Ren, C. Silberhorn, K. Banaszek, I.A.
Walmsley, to be published.


\end{thebibliography}

\end{document}